\documentstyle[preprint,amstex,amsfonts,amssymb,epsfig,aps]{revtex}
\tightenlines
\newcommand{\rv}{{\bf r}}

\newcommand{\kv}{{\bf k}}
\newcommand{\Rv}{{\bf R}}
\newcommand{\mvec}{{\bf m}}

\newcommand{\tauv}{{\boldsymbol\tau}}

\newcommand{\half}{{1\over 2}}
\newcommand{\R}{{\mathbb R}^3}
\renewcommand\eqref[1]{Eq.~(\ref{#1})}

\newcommand\cref[1]{Ref.~\cite{#1}}
\newcommand\abbref[1]{Fig.~\ref{#1}}

\newcommand\abbpref[1]{Fig.~\protect\ref{#1}}

\newcommand\rhos{\rho^\ast}
\newcommand\ms{m^\ast}
\newcommand\ts{T^\ast}
\newcommand{\erfc}{\operatorname{erfc}}

\begin{document}
\draft
\title{Crystal structures and freezing of dipolar fluids}
\author{B. Groh$^1$ and S. Dietrich$^{2,3}$}
\address{$^1$Fachbereich Physik, Bergische Universit\"at Wuppertal,
  D--42097 Wuppertal, Germany \\
   $^2$Max-Planck-Institut f\"ur Metallforschung, Heisenbergstr. 1,
   D--70569 Stuttgart, Germany \\
   $^3$Institut f\"ur Theoretische und Angewandte Physik,
   Universit\"at Stuttgart, Pfaffenwaldring 57, D--70569 Stuttgart, Germany}
\date{\today}

\maketitle

%%%%%%%%%%%%%%%%%%%%%%%%%%%%%%%%%%%%%%%%%%%%%%%%%%%%%%%%%%%%%%%%%%%%%%%%%
\begin{abstract}

We investigate the crystal structure of classical systems of spherical
particles with an embedded point dipole at $T=0$. The ferroelectric ground state energy is
calculated using generalizations of the Ewald summation technique. Due to the
reduced symmetry compared to the nonpolar case the crystals are never
strictly cubic. For the Stockmayer (i.e., Lennard-Jones plus dipolar)
interaction three phases are found upon increasing the dipole moment:
hexagonal, body-centered orthorhombic, and body-centered tetragonal. An even
richer phase diagram arises for dipolar soft spheres with a purely
repulsive inverse
power law
potential $\sim r^{-n}$. A crossover between qualitatively different
sequences of phases occurs near the exponent $n=12$. The results are
applicable to electro- and magnetorheological fluids. In addition to
the exact ground state analysis we study freezing of the Stockmayer
fluid by density-functional theory.

\end{abstract}

%%%%%%%%%%%%%%%%%%%%%%%%%%%%%%%%%%%%%%%%%%%%%%%%%%%%%%%%%%%%%%%%%%%%%%
%%%%%%%%%%%%%%%%%%%%%%%%%%%%%%%%%%%%%%%%%%%%%%%%%%%%%%%%%%%%%%%%%%%
\section{Introduction} \label{intro}

The last years have seen a revival of interest in simple dipolar
fluids, which consist of spherical particles with embedded point
dipoles, triggered by two unexpected observations: At high interaction
strengths and high densities a liquid phase with long-range ferroelectric
orientational order but without any positional
order occurs \cite{ReviewNote}; at low
densities the particles form a gas of chains behaving like living
polymers \cite{TeixeiraRev:00}. The latter effect was suggested as an
explanation for the apparent absence of gas-liquid condensation in
dipolar hard spheres \cite{vanRoij:96}, but this subject is still under discussion
\cite{Shelley:99}. Both phenomena have first been detected in computer
simulations \cite{Wei:92,Weis:92,Stevens:95b,Weis:93a,Stevens:95a},
followed by theoretical work
\cite{Letter,Paper,Klapp:97,Sear:96,Osipov:96}. This discussion
applies to electric and magnetic dipoles in complete analogy; in the
following we will use the electric language.

Knowledge about the solid phase of these systems is necessary in order
to know under which circumstances formation of the ferroelectric liquid is
preempted by freezing. This question has been tackled theoretically
within two different versions of density-functional theory. Groh and Dietrich
\cite{Solid} found a stable ferroelectric liquid for the Stockmayer
(i.e., dipolar Lennard-Jones) model if the dipole moment is high
enough. Klapp and coworkers, however, found within their approach that
the ferroelectric liquid is always metastable in comparison with the solid
both for Stockmayer and dipolar hard sphere fluids \cite{Klapp:97a,Klapp:98}. But in a
recent study they demonstrated that their result depends sensitively on
the applied approximations \cite{Klapp:00}. The only solid structures considered in these
papers \cite{Solid,Klapp:97a,Klapp:98,Klapp:00} are face-centered cubic (fcc), which is the known crystal
structure in the non-polar limit, and body-centered tetragonal (bct)
 with the special axis ratio $c/a=\sqrt{2/3}$, which had been
determined before as the ground state of dipolar hard spheres
\cite{Tao:91} and has also been observed in simulations \cite{Martin:98} and
in experiments \cite{Chen:92,Dassanayake:00} with electrorheological fluids, i.e.,
suspensions of polarizable colloidal particles.

In a recent simulation Gao and Zeng \cite{Gao:00} confirmed the
occurrence of a stable ferroelectric liquid phase for the Stockmayer
model and, in addition, determined for the first time portions of the
phase boundaries between isotropic and ferroelectric liquid and
between ferroelectric liquid and solid. Although these results support
the basic conclusion in \cref{Solid} they found that
the stable solid phase of the Stockmayer model is body-centered
orthorhombic (bco) and additionally observed a metastable distorted
hexagonal structure, possibilities that have not been taken into
account in the theoretical work
\cite{Solid,Klapp:97a,Klapp:98,Klapp:00} so far. Both bco and bct as well as an
fcc crytal with helically varying polarization direction have been
reported before in simulations of dipolar hard spheres \cite{Weis:93},
but the thermodynamically stable state could not be determined. 
Certain structures may be suppressed in simulations due to the
periodic boundary conditions if the cell shape is not flexible enough.

A simple heurestic argument why a cubic structure is not expected runs
as follows.
All these crystals are ferroelectric and hence have less symmetry than
in the nonpolar case; the point symmetries can only be reflections
at planes that
contain the polarization axis and rotations around this
axis. Therefore if, e.g., in a cubic crystal a
polarization in $[100]$ direction is switched on, the additional
introduction of a contraction along
this direction does not further reduce the remaining symmetry. Hence
generically the crystal will have an axis ratio different from unity,
i.e., it will be tetragonal. For polarization along the initial
$[110]$ and $[111]$ directions the same reasoning leads to
orthorhombic and trigonal crystals. Thus a ferroelectric solid can
never be strictly cubic, in contrast to the assumptions made in the
density-functional work described above and in similar work on the
Heisenberg fluid \cite{Lomba:98}.

Thus up to now it is quite unclear which crystal structure(s) actually are stable in these widely
used dipolar model systems. As a first step to
elucidate this region of their phase diagram in the present work we
determine the ground state structure as a function of density, dipole
strength, and softness of the isotropic interaction potential. Since
in principle an infinite number of crystal structures exists,
characterized by an increasing number of parameters with increasing
number of particles in the basis, an exhaustive search for the ground
state is not possible. However, we include a rather large class of
candidate structures, comprising among others all reasonable simple
Bravais lattices. A relatively complex phase behavior is found at
$T=0$. There is reason to expect that the ground state results remain qualitatively
valid at not too high temperatures and thus provide a valuable
starting point for a more complete analysis of the full phase diagram.

In a recent work \cite{Fernandez:00} ground state energies and
ordering temperatures for ferroelectric and antiferroelectric
arrangements of Ising dipoles on several lattices have been determined, but
no attempt has been made to optimize the lattice structure. In this
study also the presence of
isotropic interactions was not taken into account.

In our analysis we always assume a spatially homogeneous polarization throughout the
sample. In real ferroelectric materials domains will form due to the
long range nature of the dipolar interaction \cite{Domains}. The
domain structure in the {\it liquid} ferroelectric phase for a cubic
sample shape has been analyzed in detail in \cref{Simann}. This more
complicated situation is avoided in two cases: (i) a needle-shaped
sample (infinite aspect ratio), implying a vanishing depolarization
factor; (ii) cancelling of the induced surface charges by free charges
which are present in a conducting surrounding medium or as impurities
in the material itself.

After having reached a basically complete overview of the
thermodynamically stable {\it ground} states in such systems, which is
interesting in its own right, in a second step  by using
density-functional theory we return to the issue 
which solid phase the ferroelectric liquid phase the Stockmayer
fluid forms upon {\it freezing}. This closes the
aforementioned gap between the analysis of the solid phases as
considered so far in the
theoretical analyses \cite{Solid,Klapp:97a,Klapp:98,Klapp:00} and the
observation of
those types of solid structures as found in simulations \cite{Gao:00}. It consolidates the
theoretical prediction \cite{Solid} that the Stockmayer fluid can
exhibit a thermodynamically stable ferroelectric liquid phase, in
accordance with the simulation results.

%%%%%%%%%%%%%%%%%%%%%%%%%%%%%%%%%%%%%%%%%%%%%%%%%%%%%%%%%%%%%%%%%%%
\section{Models and Methods}

We study systems of spherical dipolar particles interacting via the
dipolar potential
\begin{equation} \label{wdip}
  w_{dip}(\rv)=\frac{-m^2 r^2+3({\bf m}\cdot\rv)^2}{r^5}
\end{equation}
where $\rv$ is the interparticle vector and ${\bf m}$  the dipole
moment, which is assumed to have the same orientation for all particles. In addition
there is an isotropic interaction $w_{iso}$ which is taken as either
the purely repulsive soft sphere (SS) potential
\begin{equation}
  w_{SS}(r)=4\epsilon \left(\frac{\sigma}{r}\right)^n
\end{equation}
or the Lennard-Jones (LJ) potential
\begin{equation}
  w_{LJ}(r)=4\epsilon
  \left[\left(\frac{\sigma}{r}\right)^{12}-\left(\frac{\sigma}{r}\right)^6\right].
\end{equation}
The parameters $\epsilon$ and $\sigma$ define the energy and length
scales, respectively. The ground state properties only depend on the reduced dipole
moment $\ms=(m^2/(\sigma^3\epsilon))^{1/2}$ and the reduced particle
density $\rhos=\sigma^3 N/V$ of $N$ particles in the volume $V$. Since in the SS case only the
combination $\epsilon\sigma^n$ enters the potential, the only
independent thermodynamic parameter is $x={\ms}^2/{\rhos}^{n/3-1}$. In
the LJ case the ground state energy per particle $U$ has the form
$U=\epsilon \hat E(\rhos,\ms)$ whereas for the SS case one has
$U=\epsilon{\rhos}^{n/3} \bar E({\ms}^2/{\rhos}^{n/3-1})$ where $\hat
E$ and
$\bar E$ are dimensionless scaling functions which contain all the
structure dependence. For soft spheres the optimum structure remains
the same as long as $x$ is kept constant because the prefactor
${\rhos}^{n/3}$ is the same for all structures belonging to a given
reduced 
density $\rhos$. Furthermore one can 
show that also phase coexistence densities $\rhos_i$ determined for
one value of $\ms$ can be scaled to another dipole moment ${\ms}'$
according to ${\rhos}'_i=(\ms/{\ms}')^{6/(n-3)} \rhos_i $, because
$\partial(\rhos U)/\partial\rhos=\epsilon{\rhos}^{n/3} E_1(x)$ with
another scaling function $E_1$.  Hence for
the SS model
without loss of generality we set $\ms=1$.

Explicitly the ground state energy per particle is
\begin{equation} \label{Ulatt}
  U=U_{dip}+U_{iso}=\frac{1}{2} {\sum_{\Rv,\tauv}}^{\;\prime}
  \left[w_{dip}(\Rv+\tauv)+w_{iso}(\Rv+\tauv) \right]
\end{equation}
where $\Rv$ runs over the $N_l$ lattice vectors of a Bravais lattice and
$\tauv$ over the $M$ positions of the basis particles within one unit
cell so that $N=N_l M$; the prime on the summation sign indicates that the term with
$\Rv+\tauv=0$ must be omitted. Here we implicitly have replaced a
double sum over the lattice sites by a single sum, assuming that the
average energy per particle is equal to the energy of the particle at
the origin. It is a non-trivial issue whether this assumption is justified for the long-ranged
dipolar potential. In the appendix we show that it is 
correct for a homogeneous orientational configuration in an ellipsoidal sample shape, but not, e.g., for a
parallelepiped. (But we recall that only under the conditions
mentioned in the last but one paragraph of Sec.~\ref{intro} the ground
state has spatially homogeneous orientational order as assumed in
\eqref{Ulatt}.) Straightforward numerical calculation
of these lattice sums is hampered by their slow convergence. Therefore
generalizations of the Ewald technique are employed to transform the
sums into a more rapidly converging form. The basic idea for
evaluating adroitly a general
sum $\sum_\Rv f(\Rv)$ is to write $f(\Rv)=h(\Rv)+g(\Rv)$ where
$h(\Rv)$ decays rapidly in real space and the Fourier transform
$\tilde g(\kv)=\int d^3r\,e^{-i\kv\cdot\rv} g(\rv)$ decays rapidly in reciprocal space
\cite{Smith:88}. The sum of $g$ is then evaluated in reciprocal
space using the Poisson sum formula. For the inverse power sums with
$f(\Rv)=R^{-n}$ one uses $h(R)=\Gamma(n/2,\nu^2 R^2)/(R^n \Gamma(n/2))$
where $\Gamma(a,x)$ is the incomplete Gamma function and obtains for a
Bravais lattice \cite{Nijboer:57,Smith:88}
\begin{multline}
  \sum_{\Rv\neq 0}\frac{1}{R^n}=\frac{1}{\Gamma(n/2)}\Big[
  \sum_{\Rv\neq 0} \frac{1}{R^n} \Gamma(n/2,\nu^2
  R^2)+\frac{\pi^{3/2}}{V_c} \sum_{\kv\neq 0}
  \left(\frac{k}{2}\right)^{n-3}
  \Gamma\left(\frac{3-n}{2},\frac{k^2}{4\nu^2}\right) \\
  -\frac{2}{n} \nu^n+\frac{2}{n-3}\frac{\pi^{3/2}}{V_c} \nu^{n-3} \Big].
\end{multline}
Here $\kv$ runs over the reciprocal lattice, $V_c$ is the volume of
the unit cell and the last two contributions take into account the
omissions of the terms $R=0$ and $k=0$. The parameter $\nu$ can be chosen
arbitrarily and the independence of $\nu$ of the total sum provides a convenient check
of the algorithm. In practice $\nu$ is chosen such that both sums
converge approximately with the same rate. Typically a few hundred lattice
vectors in real and reciprocal space are sufficient to obtain machine
precision ($10^{-16}$). The straightforward generalization to $M$ particles in the
basis leads to 
\begin{multline}
  U_{SS}^{(n)}=\frac{2\epsilon\sigma^n}{\Gamma(n/2)} \Big[
  {\sum_{\Rv,\tauv}}^{\;\prime}
  \frac{\Gamma(n/2,\nu^2(\Rv+\tauv)^2)}{|\Rv+\tauv|^n}
  +\frac{\pi^{3/2}}{V_c} \sum_{\kv\neq 0} \sum_\tauv
  \cos(\kv\cdot\tauv)   \left(\frac{k}{2}\right)^{n-3}
  \Gamma(\frac{3-n}{2},\frac{k^2}{4\nu^2}) \\
  -\frac{2}{n} \nu^n+N \frac{2}{n-3}\frac{\pi^{3/2}}{V_c} \nu^{n-3}
  \Big].
\end{multline}
Clearly for the LJ case $U_{iso}=U_{SS}^{(12)}-U_{SS}^{(6)}$.

An analogous expression for the dipolar lattice sum can be obtained
from the well-known result for the potential of a lattice of point
charges \cite{Slater:67}
\begin{equation} \label{udipew}
  U_{dip}=\frac{m^2}{2} \left[\sum_{\Rv\neq 0}
  \Phi_\nu(\Rv)+\frac{4\pi}{V_c} \sum_{\kv\neq 0} \frac{k_z^2}{k^2}
  \exp(-\frac{k^2}{4\nu^2}) -\frac{4 \nu^3}{3\sqrt{\pi}} \right]
\end{equation}
with
\begin{equation}
  \Phi_\nu(\Rv)=\frac{2\nu}{\sqrt{\pi} R^4} (R^2-3R_z^2-2\nu^2 R^2
  R_z^2) e^{-\nu^2 R^2}+\frac{R^2-3R_z^2}{R^5} \erfc (\nu R).
\end{equation}
Due to the slow decay of $w_{dip}$ the dipolar sum is actually only
conditionally convergent, i.e., the result is shape dependent. In the
appendix we present the derivation of \eqref{udipew} and show that it
corresponds to the case of a needle-shaped sample which is of interest
here. The corresponding generalization to lattices with a basis reads
\begin{equation}
  U_{dip}=\frac{m^2}{2} \left[ {\sum_{\Rv,\tauv}}^{\;\prime} \Phi_\nu(\Rv+\tauv)
  +\frac{4\pi}{V_c} \sum_{\kv\neq 0}\sum_\tauv \cos(\kv\cdot\tauv) \frac{k_z^2}{k^2}
  \exp(-\frac{k^2}{4\nu^2}) -\frac{4 \nu^3}{3\sqrt{\pi}} \right].
\end{equation}

If the dipoles are not oriented along the $z$ axis of the lattice but
have a general orientation $\hat\mvec$ the dipole sum has the form
$\sum_{ij} \hat m_i T_{ij} \hat m_j$ with a symmetrical matrix
$T_{ij}$ (see \eqref{wdip}). The optimum direction is then necessarily
along one of the eigenvectors of $T$ which coincide with high symmetry
lattice directions. Thus it suffices to consider only one or two
possible orientations for each lattice.

\begin{figure}
\centerline{\epsfig{file=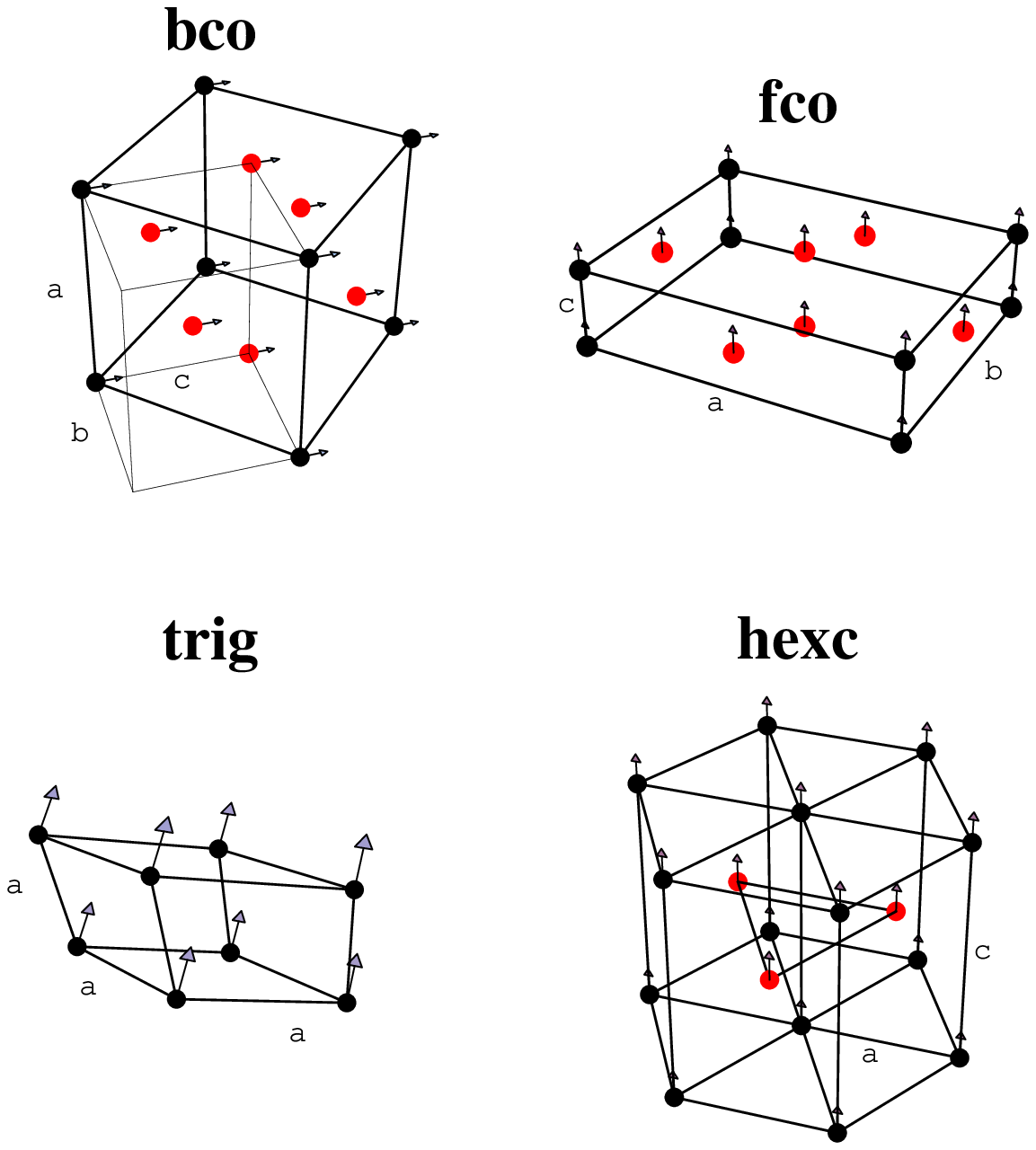,width=14cm}}
\caption{Geometries of the first four lattice structures considered in
  the search for the energy minimum. For bco the situation at the fcc-bco
  transition is shown: when a small
   dipole moment is introduced in a nonpolar fcc crystal, polarization
   along $[110]$ is preferred and the crystal contracts in this
   direction. Thick and thin lines mark the conventional unit cells
   of fcc and bco, respectively. Red particles lie on the face
  centers for fcc and fco and at half height ($z=c/2$) for hexc.}
\label{fig:lattices}
\end{figure}

\begin{figure}
\centerline{\epsfig{file=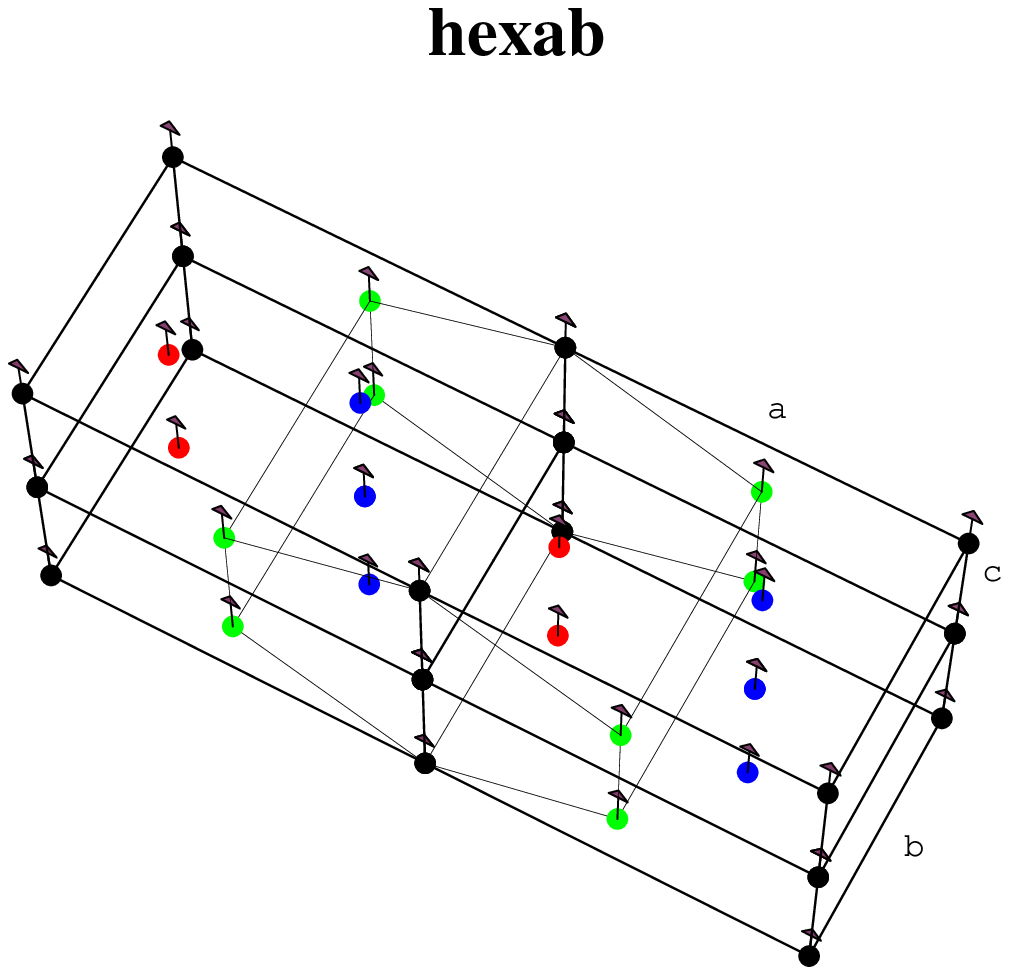,width=12cm}}
\vspace{.5cm}
\caption{The ``hexab'' crystal structure which arises from a
  distortion of a hexagonal close packed lattice polarized in a
  nearest-neighbor direction within the hexagonal plane. It is an
  orthorhombic structure with three additional basis particles shown in
  red ($\tauv_1$), green ($\tauv_2$), and blue ($\tauv_3$) with
  coordinates as given in the main text. Four orthorhombic unit
  cells are shown. The thin lines mark the hexagonal unit cell that is
  recovered for the special axis ratio $c/a=1/\protect\sqrt{3}$.}
\label{fig:monhex}
\end{figure}

The following crystal structures, displayed in
Figs.~\ref{fig:lattices} and \ref{fig:monhex}, were included in the search
for the minimum of the energy
(except for trigonal lattices the polarization is always along the $c$ axis):
\begin{enumerate}
\item body-centered orthorhombic (bco) with axis lengths $a$, $b$,
  $c$; reduces to bct for $a=b$, to fcc polarized along $[110]$ for
  $b/a=c/a=1/\sqrt{2}$ (see \abbref{fig:lattices}), and to fcc polarized
  along $[001]$ for $b/a=1$, $c/a=\sqrt{2}$.

\item face-centered orthorhombic (fco); note that for the tetragonal
  case ($b=a$) face-centered and body-centered lattices are
  equivalent.

\item trigonal (trig): three equal axes with angle $\gamma$ between
  any pair of them, polarized along $[111]$; reduces to various cubic
  lattices polarized along $[111]$ for  special values of $\gamma$.

\item hexagonal with axis lengths $a$, $c$ and a second basis
  particle at $\tauv=a/2 (1,1/\sqrt{3},c/a)$ polarized along the $c$
  axis (hexc); corresponds to hexagonal
  close packed (hcp) for $c/a=\sqrt{8/3}$.
  
\item an orthorhombic lattice with four
  basis particles at $\tauv_0=0$, $\tauv_1=(a/6,b/2,c/2)$,
  $\tauv_2=(a/2,0,c/2)$, and $\tauv_3=(2 a/3,b/2,0)$ which can be
  viewed as a distorted hcp lattice with polarization in the $ab$-plane
  (hexab); this structure was observed in \cref{Gao:00}.
\end{enumerate}
These possibilities have been chosen for the following reasons.
The structures bco, hexc, and hexab are generated by slight
distortions of the close packed fcc and hcp lattices which represent
the ground state for $m=0$. The structures fco and trig are included,
because they approach reasonable low density configurations in certain
limits, which have been observed in electrorheological fluids
\cite{Dassanayake:00}. For $\gamma\to 2\pi/3$ trig degenerates into a
hexagonal array of dipolar chains, and for $c,b\gg a$ fco develops
into a collection of parallel sheets. The only remaining Bravais lattices,
monoclinic and triclinic, are improbable and also difficult to
handle because of the larger number of free parameters. We tested a
monoclinic variation of bco and always found that the energy is minimized for a right
angle between the axes. Concerning
more complex lattices it is difficult to define a reasonable parameter
space without a physically motivated structure to start from.

%%%%%%%%%%%%%%%%%%%%%%%%%%%%%%%%%%%%%%%%%%%%%%%%%%%%%%%%%%%%%%%%%%%
\section{Results}

\subsection{Stockmayer model at $T=0$}

\begin{figure}
\centerline{\epsfig{file=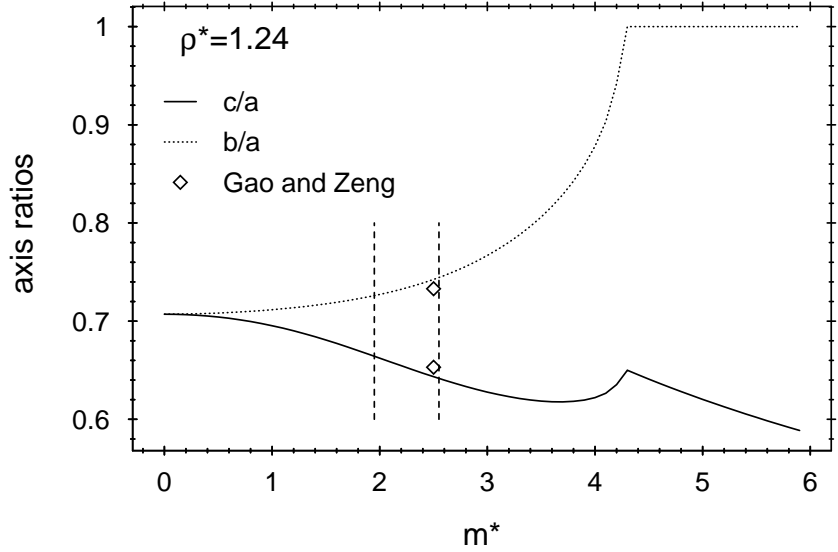,width=12cm}}
\caption{Axis ratios of the bco phase of a Stockmayer solid at $T=0$ and
  $\rhos=1.24$ as a function of the dipole moment. For $\ms=0$ one has
  $c/a=b/a=1/\protect\sqrt{2}$ corresponding to fcc. The diamonds denote
  simulation results of Gao and Zeng \protect\cite{Gao:00} for
  $\rhos=1.24$, $\ms=2.5$, and $\ts=0.7$. At $\ms_c\simeq 4.25$ a continuous transition to bct takes
  place. Note that for $\ms<1.95$ the hexc structure is slightly more stable than bco,
  while for $\ms>2.55$ gas-solid phase separation occurs. Thus for
  this density bco is stable within the indicated window and
  metastable outside.}
\label{fig:minrf}
\end{figure}

If one starts from the fcc structure of the nonpolar LJ system and
introduces a dipole moment, all orientations of the polarization have
the same energy, as long as the cubic symmetry is preserved. However,
as discussed in the introduction, the crystal actually distorts
towards bco, bct, or trig, depending on the direction of the
polarization. The numerical findings indicate that the $[110]$
direction is selected, corresponding to a bco distortion as shown in
\abbref{fig:lattices}.  Intuitively this can be understood as a
preference for the formation of chains along the polarization
direction: $[110]$ points towards the nearest neighbors so that the
particle distance along the chains is smallest in this case. Indeed,
upon increasing the dipole moment the length of the $c$ axis
decreases, reflecting the strong attractive interactions along the
chains. Figure~\ref{fig:minrf} displays the dependence of the bco axis
ratios on the reduced dipole moment $\ms$.  The result of Gao and Zeng
\cite{Gao:00} for $\rhos=1.24$, $\ms=2.5$, and $\ts=k_B T/\epsilon=0.7$ denoted by
the diamonds lies very close to the $T=0$ result. Obviously the ground
state gives a good approximation to the equilibrium state at least up
to half the triple temperature $\ts_{tr}\simeq 1.3$ \cite{Gao:00}. The
same authors also report a metastable hexab structure at $\ms=2.5$,
$\ts=0.8$, and $\rhos=1.146$  with the axis ratios $c/a=0.497$ and
$b/a=0.953$. The corresponding values  in the
ideal hcp lattice are $c/a=0.577$ and $b/a=0.942$,
respectively. At $T=0$ the hexab crystal is only metastable, too, with
$c/a=0.510$ and $b/a=0.961$. With
increasing dipole moment the
two bco axis lengths $b$ and $a$ perpendicular to $\mvec$ approach
each other  until at a critical value
$\ms_c$ a continuous transition to bct takes place
(\abbref{fig:minrf}).  After a cusp at $\ms_c$ in the bct phase $c/a$
decreases again and is much lower than the ``ideal'' value
$\sqrt{2/3}=0.816$ found for dipolar hard spheres.

\begin{figure}
\centerline{\epsfig{file=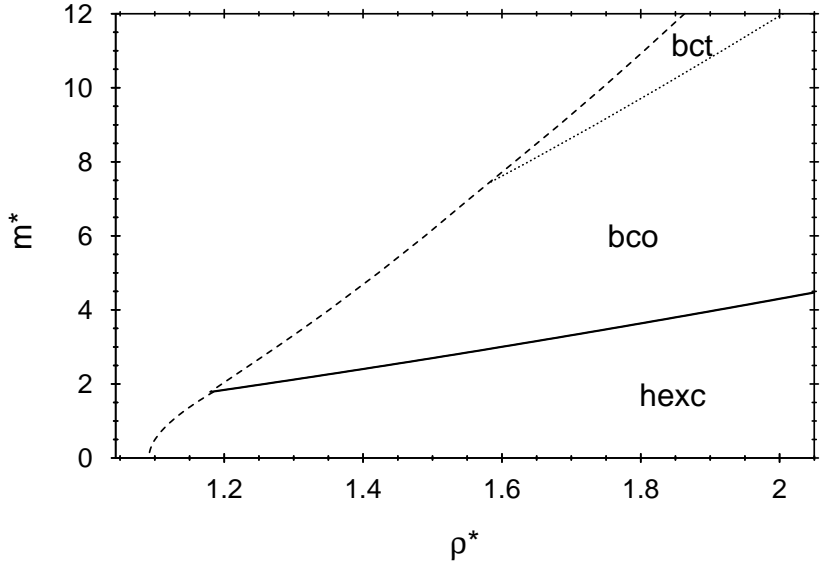,width=12cm}}
\caption{Ground state phase diagram of the Stockmayer model. The solid
  (dotted) line denotes first (second) order transitions. The dashed line
  indicates the global minimum of the energy per particle over all
  densities. The unlabeled area to the left of this line is a two-phase
  region where an infinitely diluted gas coexists with the solid. The
  maximum of the pressure $p=\rho^2 \partial U/\partial\rho$ in the
  depicted parameter range occurs in the lower right corner where
  $p\simeq 500\epsilon/\sigma^3$. Using parameter values for argon this
  corresponds to about 20$\,$GPa which is accessible in a laboratory.}
\label{fig:phasstock}
\end{figure}

The true ground state of the nonpolar LJ model, however, is hcp which has an
energy very slightly below the fcc value (by 0.01--0.02\% depending on density). This difference is only due
to second nearest neighbors because the number and distances of the 12
nearest neighbors are equal in fcc and hcp. The dipolar energy of the
hcp lattice with polarization along the $c$ axis \cite{Martin:98} is
lower than with polarization along the $a$ axis
\cite{Tao:91}. Therefore for small dipole moments a slightly
contracted hcp lattice (hexc) is preferred over bco, although the
energy differences are always very small, e.g., below 0.16\%
for $\rhos=1.24$. At larger dipole moments
bco takes over as the stable phase. Because of the smallness of these differences
entropy effects at finite temperature may easily tip the balance towards
one or the other phase. The full $T=0$ phase diagram
is shown in \abbref{fig:phasstock}. The hexc-bco transition is first
order but exhibits only a tiny density jump $\Delta\rhos\simeq 5\cdot
10^{-3}$. The dashed line $\rhos_{min}(\ms)$ marks the absolute minimum
of the energy per particle over all densities. If a system is prepared
with a density below $\rhos_{min}$ it spontaneously shrinks to this
minimum and leaves a corresponding portion of empty space. Thus the
region $\rhos<\rhos_{min}$ is a two-phase coexistence region between the
lowest energy solid and an infinitely diluted gas. For $T>0$ it
connects to  gas-solid coexistence while the liquid phase(s) appear
only at higher temperature above a triple point. 
That $\rhos_{min}$ corresponds to a phase coexistence density can also
be shown more
formally by performing a double tangent construction on the (free) energy
  density $u(\rho)=\rho U(\rho)$. For finite temperatures an entropic
term $\sim T\rho \ln\rho$ must be added for the gas phase, so that a
double tangent at $\rho_g$ and $\rho_s$ can be constructed with
\begin{equation}
  \rho_g\to 0, \qquad \rho_s\to\rho_{min}, \qquad \left.\frac{d
  u}{d\rho}\right|_{\rho_{min}}=\left.\frac{u}{\rho}\right|_{\rho_{min}},
  \qquad \text{for } T\to 0.
\end{equation}
The last equation is equivalent to the minimum condition $dU/d\rho=0$. 

Thus we conclude that at $T=0$ the Stockmayer crystal in the commonly
studied range $\ms\lesssim 3$ is either hexc or bco, but neither fcc
nor bct as assumed in various studies before.

%%%%%%%%%%%%%%%%%%%%%%%%%%%%%%%%%%%%%%%%%%%%%%%%%%%%%%%%%%%%%%%%%%%
\subsection{Stockmayer model at $T>0$} \label{SecDFT}

In this section we present the predictions of the density-functional
theory (DFT), which we applied to freezing of the Stockmayer fluid in our
previous work \cite{Solid}, when the additional crystal structures
hexc and bco are taken into account. This approach is based on a
perturbation expansion around the hard sphere solid which is treated
in the modified weighted-density approximation of Denton and Ashcroft
\cite{Denton:89}. The long-ranged isotropic and dipolar interactions
are added in such a way that a successful theory of the Lennard-Jones
fluid \cite{Curtin:86} is reproduced in the nonpolar limit. The
detailed definition of the density functional is given in \cref{Solid}
and therefore is not repeated here. 

In order to treat the hexc phase with more than one basis particle
some of the expressions given in \cref{Solid} must be
generalized; e.g., the expression for the long-range contribution to
the excess free energy turns into
\begin{equation} \label{FexLR}
  \frac{\Delta F_{\rm exc}^{(LR)}}{V}=-\frac{8\pi}{27}\rho^2 m^2
  \alpha_1^2 \left(1-6\sum_{\kv\neq 0} |S(\kv)|^2 P_2(\cos\theta_k)
  e^{-k^2/2\gamma} \frac{j_1(k\sigma)}{k\sigma} \right)
\end{equation}
with $S(\kv)=M^{-1} \sum_{\tauv} \exp(-i\kv\cdot\tauv)$ and the other
quantities as defined in \cref{Solid}. Besides the peak width
$\gamma$ and the orientational order parameters, the axis ratios now
appear as additional variables in the minimization. 

\begin{figure}
\centerline{\epsfig{file=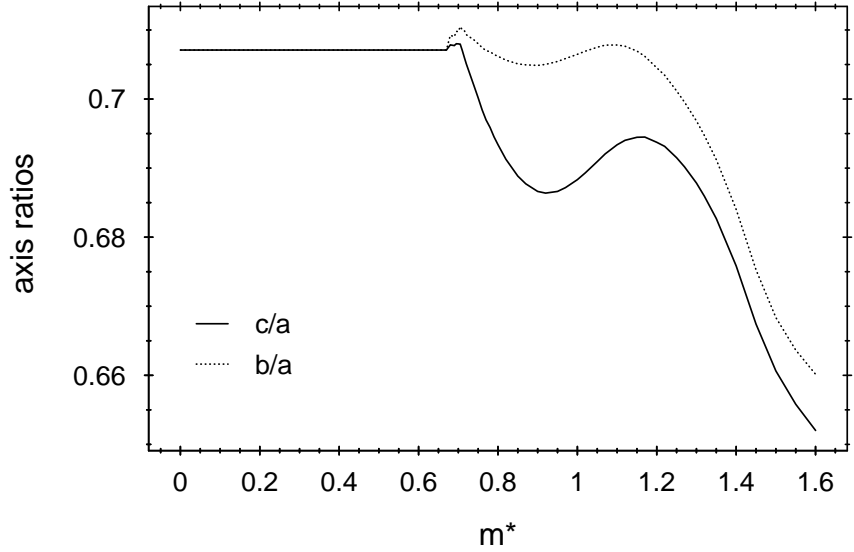,width=12cm}}
\caption{Axis ratios of the bco structure of a Stockmayer solid at $\rhos=1.24$ and
  $\ts=0.7$ calculated from density-functional theory. Ferroelectric
  order sets in for $\ms\gtrsim 0.67$. The detailed behavior near this
  point could not be clarified due to numerical problems.}
\label{fig:mindft}
\end{figure}

Figure~\ref{fig:mindft} shows the variation of the bco axis ratios as
function of the dipole moment at fixed temperature and density. For
low $\ms$ the particles in the crystal are orientationally disordered
and both axis ratios are equal to $1/\sqrt{2}$ corresponding to an fcc
lattice. When ferroelectric order sets in both axis ratios eventually
decrease but exhibit a peculiar minimum and maximum in an intermediate
range of values for $\ms$. The contraction along the polarization
direction ($c$ axis) is in accordance with the ground state result. On
the other hand, in contrast to the DFT results, both at $T=0$ and in
the simulations the value of $b/a$ is larger than $1/\sqrt{2}$ and
increases with $\ms$ (compare \abbref{fig:minrf}). A quantitative
comparison with the simulation results for $\ms=2.5$ is not possible
because such high dipole moments could not be reached in the theory.
(For large values of $\ms$ the crystalline density peaks become very narrow which leads to
convergence problems for example in \eqref{FexLR}.) In DFT usually
both axis ratios decrease with increasing dipole moment, decreasing
temperature, or decreasing density.

\begin{figure}
\centerline{\epsfig{file=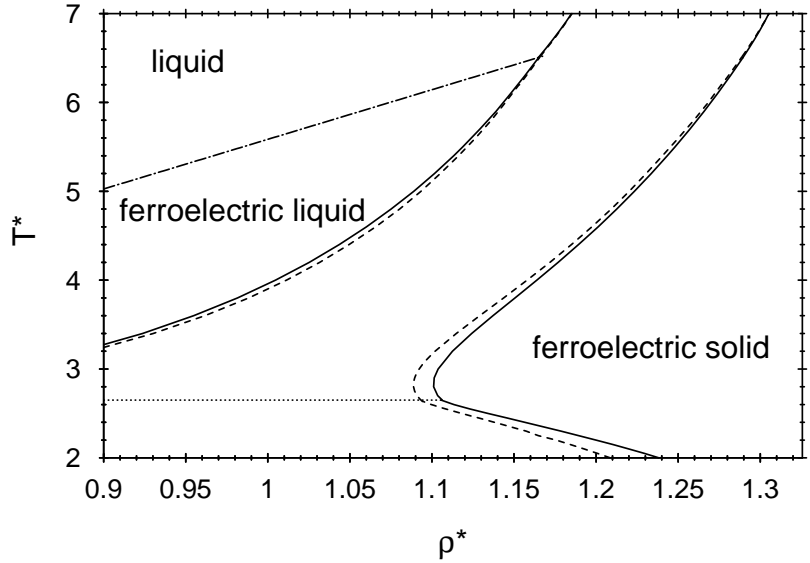,width=12cm}}
\caption{Phase diagram of the Stockmayer fluid as calculated from
  density-functional theory for $\ms=2$. The ferroelectric solid has hexc
  structure. The dotted line indicates the gas--ferroelectric
  liquid--ferroelectric solid triple temperature. The
  dashed lines are the corresponding phase boundaries if an fcc
  solid is assumed. The unlabeled areas are two-phase regions.}
\label{fig:phas2}
\end{figure}

For $\ms=2$ hexc turns out to be the stable crystal structure at all
temperatures. In \abbref{fig:phas2} we show the calculated
ferroelectric liquid--ferroelectric solid and gas--ferroelectric solid
transition densities. For comparison our previous data assuming an fcc
structure are displayed, too. The shift due to the new crystal
structure is relatively small and has no impact on the occurrence of
the ferroelectric liquid phase as such. The axis ratio $c/a$ always lies below
the ideal hcp value $\sqrt{8/3}=1.633$; it varies between 1.58 and
1.62 along the parts of the coexistence lines shown in
\abbref{fig:phas2}. Surprisingly it decreases with increasing density
while the opposite trend is observed in the ground state. For the
lower dipole moment $\ms=1$ the free energy differences between hexc,
bco, and fcc are smaller than the numerical accuracy so that with the
present tools it is not
possible to decide which phase is the stable one. In any case, however, the loci of
the phase boundaries remain practically the same as calculated in
\cref{Solid}. Thus we conclude that our DFT prediction of the
occurrence of  a stable ferroelectric liquid phase within a certain
parameter range is not altered by
taking into account more possibilities for the crystal structure and
thus is in agreement with the findings of the simulations.

%%%%%%%%%%%%%%%%%%%%%%%%%%%%%%%%%%%%%%%%%%%%%%%%%%%%%%%%%%%%%%%%%%%%%
\subsection{Dipolar soft spheres at $T=0$}

\begin{figure}
\centerline{\epsfig{file=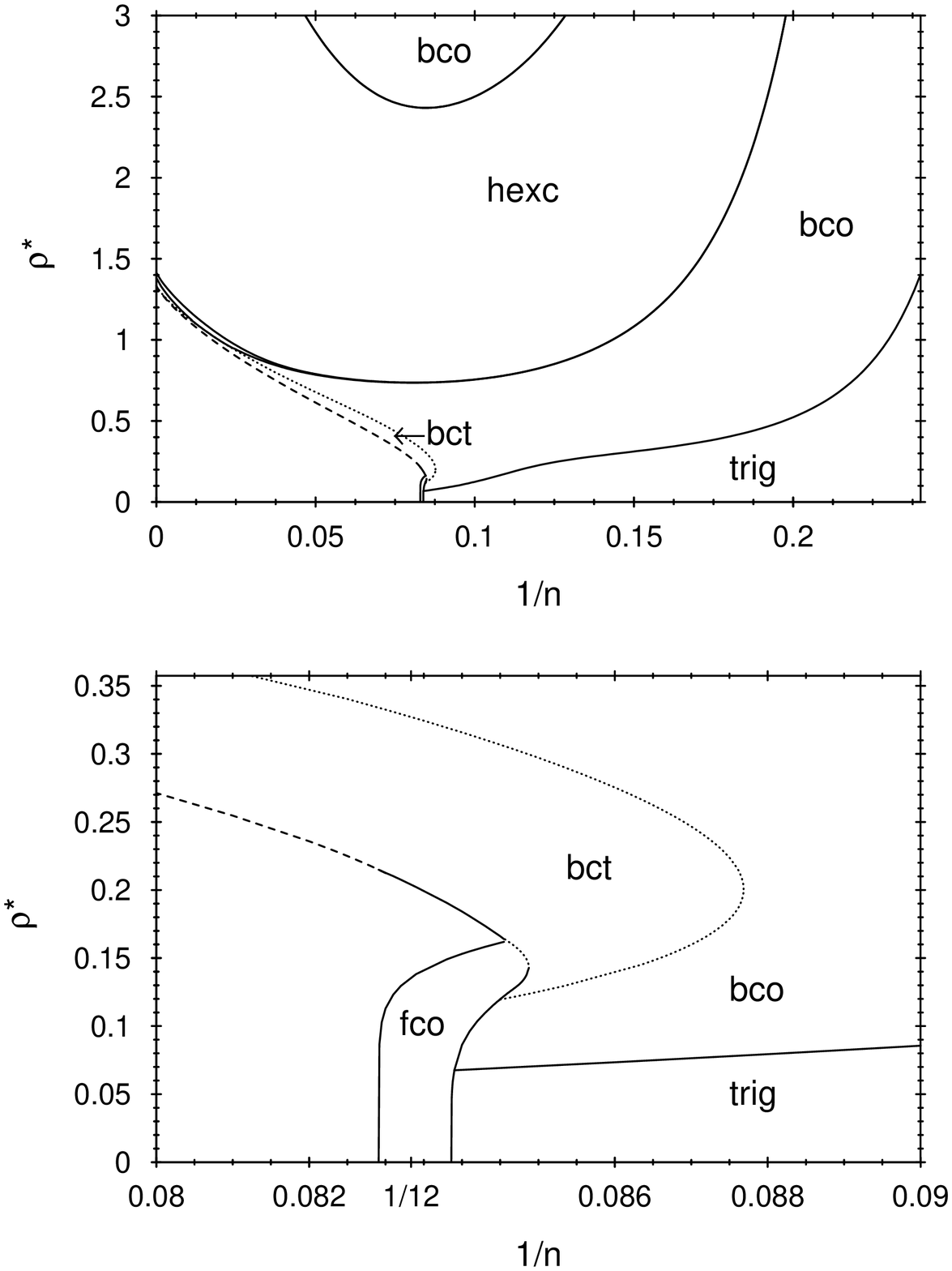,width=10cm}}
\centerline{\epsfig{file=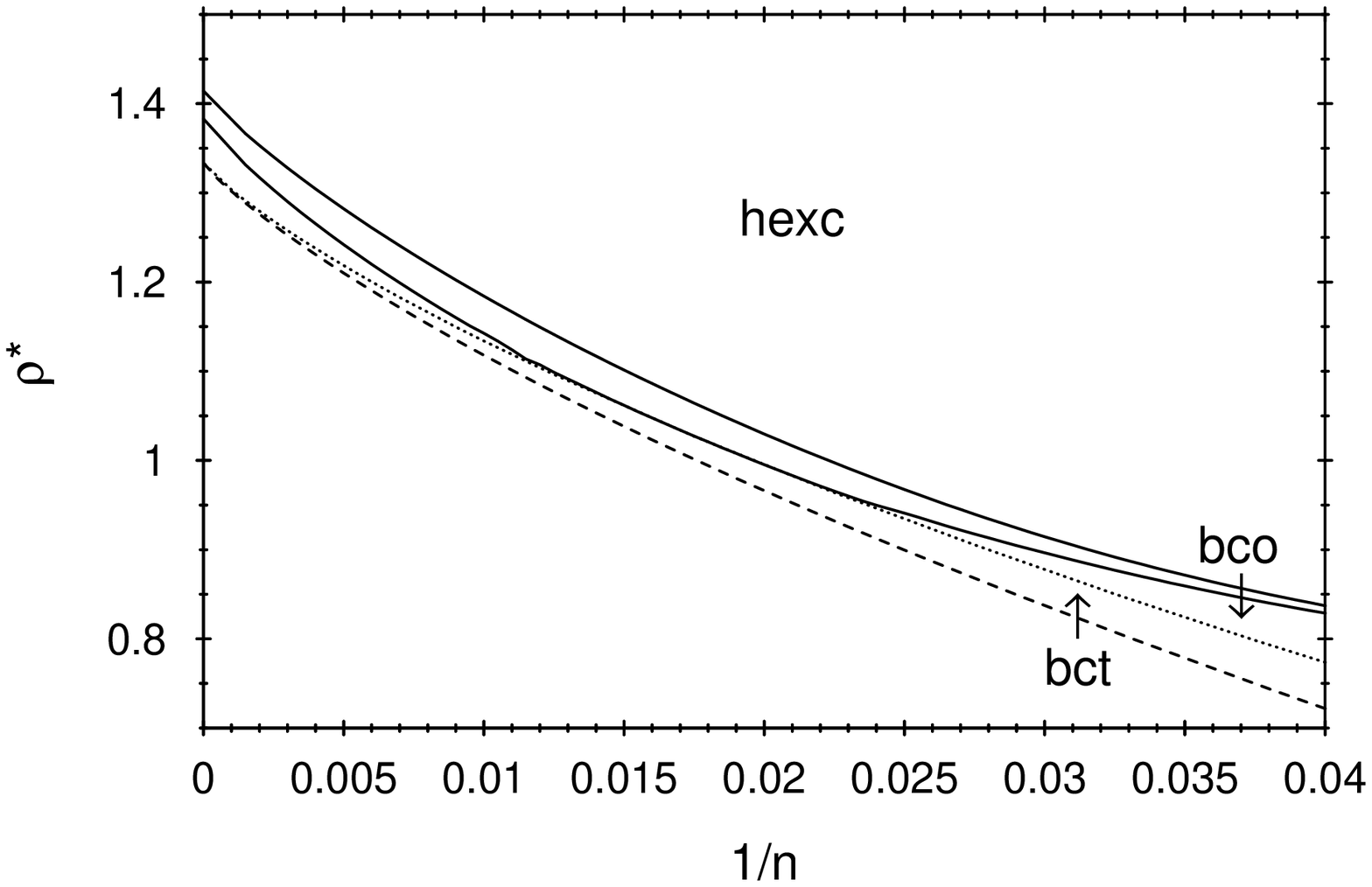,width=10cm}}
\caption{Ground state phase diagram of the dipolar soft sphere model
  for $\ms=1$. The
  meaning of the line styles is the same as in \abbpref{fig:phasstock}. The
  abbreviations for the various crystal structures are explained in the main
  text. The unlabeled areas are two-phase regions. The plots in the
  middle and at the bottom are
  magnifications of the regions around $n=12$ and near the hard sphere
  limit $n=\infty$, respectively. The phase diagram for
  other dipole moments $\ms$ can be inferred from the one given here
  by rescaling the reduced density according to
  $\rhos/{\ms}^{6/(n-3)}$. In the limit $\rhos\to\infty$ the upper bco phase turns into fcc continuously.}
\label{fig:phasss}
\end{figure}

In this case fcc is more stable than hcp for $\ms=0$, or,
equivalently, for $\ms\neq 0$ and $\rhos\to\infty$. For  large exponents $n$ with
decreasing density qualitatively
the same sequence of transitions fcc--bco--bct occurs as discussed for
the Stockmayer system. But as shown by the phase diagram in
\abbref{fig:phasss} within the bco range hexc is stable in an
intermediate range. Also in contrast to the Stockmayer model the vapor
phase can coexist only with the bct solid. 

For $n\to\infty$ one recovers the
limit of dipolar hard spheres. Here the lowest energy state is the
``ideal'' bct with $c/a=\sqrt{2/3}$ at $\rhos=4/3$ \cite{Tao:91} which
coexists with an infinitely diluted state at $\rhos=0$. For higher
densities close packed bco structures with
\begin{equation}
  \frac{c}{a}=\left(\frac{3}{8}{\rhos}^2-\rhos
 \sqrt{\frac{9}{64}{\rhos}^2-\frac{1}{4}} \right)^{1/2}, \qquad
 \frac{b}{a}=\frac{2}{\rhos} \left(\frac{c}{a}\right)^2
\end{equation}
occur in which each particle has ten nearest neighbors at
distance $\sigma$. At $\rhos=1.383$ a two-phase bco--hexc region
starts that extends up to the maximum possible density
$\rhos=\sqrt{2}$ where the ideal hcp lattice with $c/a=\sqrt{8/3}$ is
stable. For $n\to\infty$ the values of the axis ratios
along the various phase boundaries in \abbref{fig:phasss} converge towards the aforementioned hard
sphere values.

A quite different behavior occurs for small exponents $n$. Here the
isotropic repulsion cannot be overcome by the dipolar attraction in a
three-dimensional structure so that the energy is lowest at
$\rho=0$ which means that the solid phase trig remains stable down to
arbitrarily low densities without encoutering a gas-solid phase
separation. 

\begin{figure}
\centerline{\epsfig{file=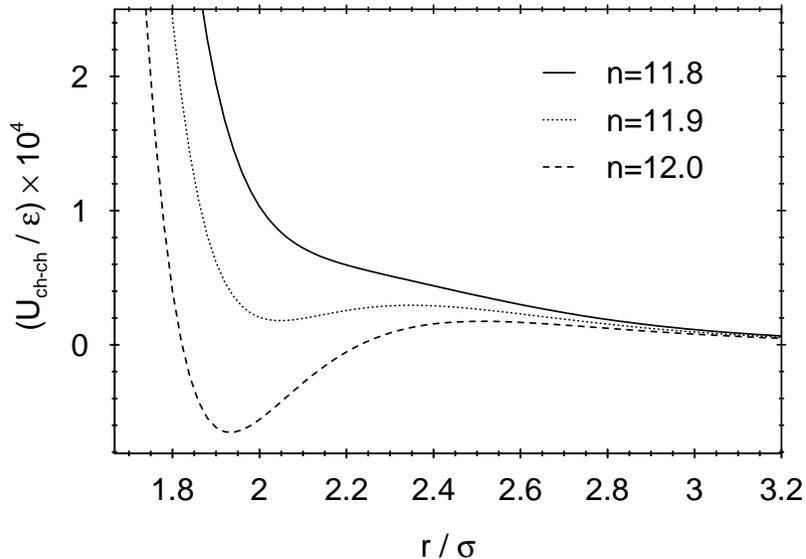,width=12cm}}
\caption{Interaction potential between parallel chains of dipolar soft
  spheres, with a longitudinal shift of half the intrachain particle
  distance $a$, for different values of the exponent $n$ as a function
  of the chain separation $r$. The plot is for $\ms=1$; corresponding
  curves for other dipole moments can be obtained by rescaling the
  distance with ${\ms}^{-2/(n-3)}$ and the energy with ${\ms}^{2
  n/(n-3)}$.}
\label{fig:uchch}
\end{figure}

A semi-quantitative understanding of this effect can be
obtained by considering arrangements of parallel chains, to which all
crystal structures degenerate for $\rho\to 0$. The intrachain energy
per particle of a soft sphere chain is \cite{Tao:91}
\begin{equation}
  U_{ch}=-2 \zeta(3) \frac{m^2}{a^3}+4 \epsilon
  \left(\frac{\sigma}{a}\right)^n \zeta(n)
\end{equation}
where $\zeta(n)=\sum_{k=1}^\infty k^{-n}$ is the Riemann zeta function
and $a$ the particle distance. The equilibrium distance follows by
minimization:
\begin{equation}
  \frac{a_{eq}}{\sigma}=\left(\frac{2 n \zeta(n)}{3\zeta(3)
  {\ms}^2}\right)^{\frac{1}{n-3}}.
\end{equation}
The purely dipolar interaction energy between two parallel chains with 
distance $r$ and longitudinal offset $z$ was calculated by Tao and
Sun \cite{Tao:91}:
\begin{equation}
\begin{split}
  U_{\rm ch-ch}^{dip} &=\frac{16\pi^2 m^2}{a^3} \sum_{k=1}^\infty
  K_0\left(\frac{2\pi k
  r}{a}\right) \cos\left(\frac{2\pi k z}{a}\right) \\
  &\simeq \frac{8\pi^2 m^2}{a^3} \sqrt{\frac{a}{r}} e^{-2\pi r/a} \cos\left(\frac{2\pi k z}{a}\right) 
\end{split}
\end{equation}
where $K_\nu$ denotes modified Bessel functions. Using similar methods
one finds for the isotropic contribution of a single power-law repulsion
\begin{equation}
  U_{\rm ch-ch}^{iso} = \frac{4\epsilon\sigma^n}{a}
  \frac{\sqrt{\pi}}{\Gamma(n/2)}
  \left[ \frac{\Gamma(\frac{n-1}{2})}{r^{n-1}}+4 \sum_{k=1}^\infty
  \left(\frac{\pi}{r a}\right)^{\frac{n-1}{2}} K_{(1-n)/2}\left(\frac{2\pi k
  r}{a}\right) \cos\left(\frac{2\pi k z}{a}\right)\right].
\end{equation}
All but the first term decay exponentially for $r\to\infty$. The total
chain-chain interaction $U_{\rm ch-ch}=U_{\rm ch-ch}^{dip}+U_{\rm ch-ch}^{iso}$, evaluated at $a=a_{eq}$ and
$z=a/2$ behaves qualitatively different for large and small exponents
$n$, as shown in \abbref{fig:uchch}. For large $n$ it exhibits a minimum with $U_{\rm ch-ch}<0$ at
intermediate $r$ followed by a maximum and eventually an algebraic
decay for $r\to\infty$, whereas for small $n$ it is repulsive and
monotonously decreasing for all distances $r$. The minimum reaches
zero at $n=11.93$ and disappears at $n=11.89$. Thus the most
frequently used
value 12 for the exponent, mainly chosen for historical reasons, just marks the crossover
between attractive and repulsive soft dipolar chains which is also reflected
in the phase diagram. Near $n=12$ a low density fco phase appears. In
this phase for $\rho\to 0$ the particles form parallel sheets, due to
the small attraction between chains (see \abbref{fig:uchch}). For
slightly smaller values of $n$ the
ground state becomes trigonal with opening angle $\gamma\to 2\pi/3$
for $\rho\to 0$, i.e., a hexagonal lattice of chains with relative longitudinal
shifts $\pm a/3$ between nearest neighbors. Using the equations above
one can verify that this limiting structure is more favorable than a
hexagonal arrangement with shifts $0$ and $a/2$ which can be obtained
from a bco lattice with $b/a=\sqrt{3}$. In the same region the bct
phase disappears so that the transition sequence upon increasing
density becomes trig--bco--hexc--bco--fcc. 

In both models the hexab phase turns out to be metastable for all parameters. All solid
lines are first order transitions. Except for the transitions bco-hexc near $n=\infty$
and fco-bct near $n=12$ the density gaps are always very small. Dashed
lines denote the continuous bco-bct and fco-bct transitions.

%%%%%%%%%%%%%%%%%%%%%%%%%%%%%%%%%%%%%%%%%%%%%%%%%%%%%%%%%%%%%%%%%%%%%%
\section{Discussion}

Even at $T=0$ the investigated model systems show a rich phase
behavior with a variety of solid-solid phase transitions as function
of dipole moment, density, and softness of the repulsion (see
Figs.~\ref{fig:phasstock} and \ref{fig:phasss}). Concerning the
experimental relevance of these phenomena molecular dipolar fluids are
natural first candidates. However, for them quantitative comparison is
impeded by the fact that typically such particles exhibit additional
short-ranged steric anisotropies which become important at freezing
densities \cite{Gay:98}. Such a sensitive dependence of the phase
behavior on the details of the repulsive part of the interaction
potential is supported by our findings concerning the decay exponent
$n$ (see \abbref{fig:phasss}). The closest effective realization of
our models are colloidal suspensions of monodisperse spherical
particles. The dipole moment can either be a permant magnetic moment
as in ferrofluids \cite{Rosensweig:85} or an induced electric or
magnetic moment as in electrorheological (ER) or magnetorheological
fluids \cite{ER:96}. At present it is still difficult to prepare
stable ferrofluids at sufficiently high densities. However, recently
the occurrence of gas--liquid--solid phase transitions in nearly
monodisperse solutions of maghemite $\gamma$-Fe$_2$O$_3$ nanoparticles
in water has been reported \cite{Dubois:00}. In this ionic ferrofluid
the effective isotropic interaction between the particles can be tuned
by changing the screened Coulomb interaction via adding salt. This
kind of intervention into the interaction potential is very
interesting since our analysis demonstrates that, as mentioned above,
the occurrence of different solid phases depends sensitively on the
details of the isotropic repulsion.
 
The electrostatic energy of an arrangement of
polarizable ER spheres with radius $a$ and dielectric constant $\epsilon_P$
in a solvent of dielectric constant $\epsilon_F$ and an external field $E$ is
\cite{Tao:91}
\begin{equation} \label{UER}
  U_{ER}=-\frac{\alpha \epsilon_F a^3 E^2}{2(1+2\alpha a^3 U^\ast_{dip})}
\end{equation}
with $\alpha=(\epsilon_P-\epsilon_F)/(\epsilon_F+2\epsilon_P)$ and $U^\ast_{dip}=U_{dip}/m^2$
the corresponding reduced energy for permanent dipole moments $m$. For small $\alpha$
  ($|\alpha|<1/2$ by definition, $\alpha\simeq0.3$ has been estimated for a silicon oil ER
fluid \cite{Lemaire:92}, while $\alpha\gtrsim -1/2$ for water based
fluids) \eqref{UER} can be expanded and the structure dependent terms take on the
same form as for permanent dipoles with an effective moment $m_{\rm
  eff}^2=\alpha^2 \epsilon_F a^6 E^2$ so that the calculated phase diagrams
apply without changes. Moreover with typical values \cite{Dassanayake:00}
$E=1 {\rm kV/mm}$ and $a=0.5\mu{\rm m}$ the dipolar energy at contact
$m_{\rm eff}^2/(2 a)^3$ is larger than the thermal energy $k_B T$ at
room temperature by three orders of
magnitude, justifying the use of our ground state analysis. Concerning
the effective isotropic
interactions, dispersion forces as modelled by the LJ potential are present
in ER fluids \cite{Woestman:93} but usually neglegibly small
\cite{Marshall:89}. Steric repulsion at small distances is achieved by
polymer coating. The length and density of polymers determines the softness of the
repulsion although it will be difficult to reproduce a power law
dependence as considered theoretically above. Thus chemical tayloring of
the particle surface represents an option to produce softly repulsive
potentials which differ from the hard sphere behavior usually assumed
in ER models.
We expect that our calculated
phase diagrams at least qualitatively reflect the behavior of such ER
fluids. While the subtle issue of the relative stability of the fcc
and hcp phases and the corresponding bco-hexc
transitions are probably masked by neglected effects such as higher
multipoles, many-particle interactions, and polydispersity, the phase sequence
fcc--bco--bct upon increasing field strength should be insensitive to these details.

At small field strengths or if the colloidal particles are covered by
nonmagnetic or hardly polarizable spherical shells the dipolar energy
at contact becomes comparable to $k_B T$. In Subsec.~\ref{SecDFT} a
previously developed density-functional theory has been applied to  calculate  the
phase diagram including the temperature as a relevant thermodynamic
variable. The overall effect of considering crystal structures
different from fcc on the position of the phase boundaries is rather
small. The occurrence of a stable ferroelectric liquid phase within a
certain parameter range is confirmed and in agreement with
corresponding conclusions based on simulation data. Nonetheless it is
conceivable that some trends in the behavior of the axis ratios, such
as the density dependence of $c/a$ in the hexc phase and the value of
$b/a$ in the bco phase, are not correctly described by DFT, which contains a number of
uncontrolled approximations. However, at present no better theory for the
quite demanding problem of  freezing of dipolar fluids is
available.

%%%%%%%%%%%%%%%%%%%%%%%%%%%%%%%%%%%%%%%%%%%%%%%%%%%%%%%%%%%%%%%%%%%%%%
\begin{appendix}

\section*{The dipolar lattice sum}

The dipolar sum is only conditionally convergent, which means that its value
depends on the order of summation, or, equivalently, on the sample shape. To
clarify this often overlooked difficulty, here we explicitly perform the
thermodynamic limit starting from finite lattices and letting the sample size
diverge for fixed shape. We consider rotational ellipsoids with axis
lengths $k L$ along the $z$ direction and $L$ along the $x$ and $y$ directions.

\subsection{Reduction to a single lattice sum}

The dipolar energy per particle is
\begin{equation} \label{udipL}
  U_{dip}^{L}=\frac{1}{2 N} \sum_{\Rv} \sum_{\Rv'\neq\Rv} w_{dip}(\Rv-\Rv')
  =\half \sum_{\Rv_{12}\neq 0} h\left(\frac{\Rv_{12}}{L}\right) w_{dip}(\Rv_{12})
\end{equation}
where the superscript $L$ refers to a finite system and the lattice vectors
$\Rv$ and $\Rv'$ run over the $N$ sites inside the sample. In the second form
one summation has been carried out so that $\Rv_{12}$ runs over a sample of doubled size,
and the function $h$ counts the number of occurrences of a given
interparticle vector $\Rv_{12}$. If for large systems the discrete
array of sites is approximated by a uniform distribution of the same
density, the function $h$ is given by the ratio of the intersection volume of
two ellipsoids shifted by $\Rv_{12}$ relative to each other and the volume of
one ellipsoid. The explicit result derived in \cref{Paper} is
\begin{align}
  h\left(\theta,\frac{R_{12}}{L}\right) &= 1+h_1(\theta) \frac{R_{12}}{L}+h_3(\theta)
  \left(\frac{R_{12}}{L}\right)^3 \\
\intertext{with}
  h_1(\theta) &= -\frac{3}{2}
  \left(\sin^2\theta+\frac{1}{k^2}\cos^2\theta\right)^{1/2} \\
\intertext{and}
  h_3(\theta) &= \frac{1}{2}
  \left(\sin^2\theta+\frac{1}{k^2}\cos^2\theta\right)^{3/2} 
\end{align}
where $\theta$ is the angle between $\Rv_{12}$ and the $z$ axis. For
rapidly decaying potentials in the thermodynamic limit $L\to\infty$
$h$ can be replaced by 1 because the difference $h-1$ becomes
appreciable only for large values of $R_{12}$ comparable with $L$
which, however, have a small weigth in \eqref{udipL} due to the
vanishing of $w(R_{12})$. But it is unclear whether this line of
argument also holds for the slowly decaying dipolar potential. In
order to check this we choose
a cutoff radius $R_c$ beyond which one may approximate the summation in
\eqref{udipL} by an integral. In this limit the terms with $R_{12}>R_c$ become
\begin{multline}
  2\pi\rho\int_{-1}^1 d\cos\theta \int_{R_c}^{2 L g(\theta)} dr r^2
  \frac{m^2}{r^3} P_2(\cos\theta) h(\theta,\frac{r}{L}) \\
  =2\pi \rho m^2 \int_{-1}^1 d\cos\theta P_2(\cos\theta) \left[\ln
  g(\theta)+h_1(\theta) g(\theta)+\frac{1}{3} h_3(\theta) g(\theta)^3
  \right]+O(\frac{1}{L}).
\end{multline}
Here $\rho$ is the density of dipoles, $P_2(x)=(3x^2-1)/2$ is the second
Legendre polynomial, and the function
$g(\theta)=(\sin^2\theta+\cos^2\theta/k^2)^{-1/2}$ parametrizes the surface
of the ellipsoid \cite{Paper}. Due to the specific forms of the functions
$h_i$ and $g$ the last two terms vanish so that indeed one obtains the same
result if $h=1$ is set from the beginning. The same is obviously true for the
contributions from $R_{12}<R_c$ for $L\to\infty$. These considerations justify that the
average energy per particle may be replaced by the energy of the central
particle. We emphasize that this argument hinges on the  ellipsoidal
shape of the sample. For example, by explicit calculations one can show that the replacement is {\it not} correct
for a parallelepiped with general aspect ratio.

\subsection{Shape dependence in the Ewald method}

The Ewald form of the dipolar lattice sum is often used without any
discussion of the shape dependence of the original sum
\cite{Adams:76,KittelEwald,Gay:98}. In the following we show that it
actually corresponds to a specific choice of the sample shape and we
derive the corrections that must be applied for other shapes. To this
end the
dipolar sum is constructed by superposition of two slightly shifted
opposite point charge lattices. Hence we first recapitulate the
derivation of the corresponding Ewald sum for the electrostatic
potential $\phi(\rv)$ of a finite Bravais lattice of positive unit
point charges plus an opposite uniform background charge.  The Ewald
method proceeds by rearranging the charge density $\rho(\rv)$ into two
contributions, employing a suitable addition of zero: first
$\rho_1(\rv)$ corresponding to a lattice of positive Gaussian charge
distributions plus the negative background, second $\rho_2(\rv)$
corresponding to a lattice of negative Gaussians plus the positive
point charges:
\begin{align}
  \rho(\rv) &=\rho_{1}(\rv)+\rho_2(\rv)=\sum_{\Rv}
  [\rho_1^{(0)}(\rv-\Rv)+\rho_2^{(0)}(\rv-\Rv)] \\
\intertext{where}
  \rho_1^{(0)}(\rv) &=\left(\frac{\nu^2}{\pi}\right)^{3/2} e^{-\nu^2
  r^2}-\frac{1}{V_c} \Theta_c(\rv) \\
\intertext{and}
 \rho_2^{(0)}(\rv) &=\delta(\rv)-\left(\frac{\nu^2}{\pi}\right)^{3/2} e^{-\nu^2
  r^2}
\end{align}
with $\Theta_c(\rv)=1$ if $\rv$ is in the unit cell around the origin and
$\Theta_c(\rv)=0$ otherwise. The Fourier transform 
\begin{equation}
\tilde \rho_1(\kv)=\int_{\R} d^3r e^{-i \kv\cdot\rv} \rho_1(\rv)
\end{equation}
of the first contribution
is
\begin{equation} \label{rho1k}
  \tilde\rho_1^L(\kv)=\tilde\rho_1^{(0)}(\kv) {\sum_\Rv}' e^{-i \kv\cdot\Rv}.
\end{equation}
The $L$ dependence arises via the summation over the finite number of
lattice vectors, indicated by the prime on the summation sign.
The sum in \eqref{rho1k} is strongly peaked around the reciprocal lattice vectors
$\bf G$
for large systems and approaches $(2\pi)^3/V_c \sum_{\bf G} \delta(\kv-{\bf G})$
in the thermodynamic limit. The corresponding electrostatic potential is
\begin{equation} \label{phi1L}
  \phi_1^L(\rv)=\frac{4\pi}{(2\pi)^3} \int_{\R} d^3 k\, e^{i\kv\cdot\rv} \tilde\rho_1^L(\kv)/k^2.
\end{equation}
If the unit cell is chosen as the volume spanned by the basis vectors
and as
centered around $\rv=0$ one finds $\tilde\Theta_c({\bf G})=0$ for all ${\bf
  G}\neq 0$ and $\tilde\rho_1^{(0)}(k=0)=0$ due to local charge neutrality.
For the second contribution $\rho_2(\rv)$ the potential is easily obtained by integration
of the Poisson equation in spherical coordinates yielding
\begin{equation}
  \phi_2^L={\sum_\Rv}' \frac{\erfc(\nu|\rv-\Rv|)}{|\rv-\Rv|}.
\end{equation}
For $L\to\infty$ the total potential is \cite{Slater:67}
\begin{equation} \label{phiew}
  \phi(\rv)=\frac{4\pi}{V_c} \sum_{{\bf G}\neq 0} \frac{1}{G^2} e^{i{\bf
  G}\cdot\rv} \exp(-\frac{G^2}{4\nu^2}) +\sum_{\Rv}
  \frac{\erfc(\nu|\rv-\Rv|)}{|\rv-\Rv|}+C(\nu)
\end{equation}
where $C(\nu)=-\pi/(\nu^2 V_c)$ is a constant stemming from the small $k$ contributions in
\eqref{phi1L}. In this form both sums
are rapidly converging and $\phi(\rv)$ is independent of $\nu$.

The potential of a corresponding arrangement of dipoles is now obtained from
the superposition of two slightly shifted point charge lattices with opposite
signs:
\begin{equation}
  \phi_{dip}^L(\rv)=\lim_{d\to 0} \frac{m}{d} \left[ \hat\phi^L(\rv-{\bf
  d}/2)-\hat\phi^L(\rv+{\bf d}/2)+\phi^L_{uni}(\rv-{\bf
  d}/2)-\phi^L_{uni}(\rv+{\bf d}/2) \right], \qquad {\bf d}=d{\hat\mvec}.
\end{equation}
The potentials $\phi_{uni}^L$ of the uniform background charges, which do not
cancel each other in this limit, have been subtracted. Furthermore the hats
indicate that the Coulomb potential $1/r$ must also be subtracted, because
the field of the dipole at the origin must not be included. The energy of the
central dipole follows as
\begin{equation}
  U_{dip}^L=\frac{m}{2} \frac{\partial}{\partial z} \left. \phi_{dip}^L
  \right|_{\rv=0} =-\frac{m^2}{2} \frac{\partial^2}{\partial z^2} \left[
  \hat\phi^L(\rv)+\phi_{uni}^L(\rv) \right]_{\rv=0}
\end{equation}
where the limit $d\to 0$ generates the second derivative. The first term reproduces \eqref{udipew} for $L\to\infty$ (see
\eqref{phiew}). The generalization to lattices with basis follows from
summing the contributions from each sublattice. The second term is calculated using
\begin{equation}
\begin{split}
 \phi_{uni}^L(\rv=(0,0,r)) &= \rho \int d^3r' \frac{1}{|\rv-\rv'|} \\
  &= 2\pi\rho \int_{-1}^1 d\cos\theta \int_0^{2 L g(\theta)} dr' {r'}^2
  (r^2+{r'}^2-2r r'\cos\theta)^{-1/2}.
\end{split}
\end{equation}
By performing the inner integration and expanding in terms of $r$ one finds
\begin{equation}
  \frac{\partial^2}{\partial z^2}\left. \phi_{uni}^L(\rv)\right|_{\rv=0} =
  -4\pi\rho \left(\frac{1}{3} +\int_{-1}^1 d\cos\theta P_2(\cos\theta) \ln
    g(\theta) \right)=-4\pi\rho D(k).
\end{equation}
The expression in brackets is the depolarization factor $D(k)$ of an
ellipsoid with aspect ratio $k$ \cite{Paper}. All other terms are independent
of $k$ for $L\to\infty$. Thus we have derived an explicit result for the
shape dependence of the dipolar energy which is the same as for the
homogeneous dipole density studied in \cref{Paper}. Since $D(k\to\infty)=0$
\eqref{udipew} is correct for a needle-shaped sample. In other
sample shapes the dipolar energy can be lowered by domain formation.

\end{appendix}
%%%%%%%%%%%%%%%%%%%%%%%%%%%%%%%%%%%%%%%%%%%%%%%%%%%%%%%%%%%%%%%%%%%%%%
%%%%%%%%%%%%%%%%%%%%%%%%%%%%%%%%%%%%%%%%%%%%%%%%%%%%%%%%%%%%%%%%%%%%%%
%\bibliographystyle{/home/groh/tex/papers/prsty}
%\bibliography{/home/groh/tex/bib/mystuff,/home/groh/tex/bib/dipole,/home/groh/tex/bib/bifsolid}

\begin{thebibliography}{10}

\bibitem{ReviewNote}
B. Groh and S. Dietrich,  in {\em New Approaches to Problems in Liquid State
  Theory}, Vol.~529 of {\em NATO Science Series}, edited by C. Caccamo, J.-P.
  Hansen, and G. Stell (Kluwer, Dordrecht, 1999), pp.\ 173--196, and references
  therein.

\bibitem{TeixeiraRev:00}
P. Teixeira, J. Tavares, and M.~T. da~Gama, J. Phys.: Condens. Matter {\bf 12},
   411  (2000).

\bibitem{vanRoij:96}
R. van Roij, Phys. Rev. Lett. {\bf 76},  3348  (1996).

\bibitem{Shelley:99}
J. Shelley, G. Patey, D. Levesque, and J. Weis, Phys. Rev. E {\bf 59},  3065
  (1999).

\bibitem{Wei:92}
D. Wei and G. Patey, Phys. Rev. Lett. {\bf 68},  2043  (1992).

\bibitem{Weis:92}
J. Weis, D. Levesque, and G. Zarragoicoechea, Phys. Rev. Lett. {\bf 69},  913
  (1992).

\bibitem{Stevens:95b}
M. Stevens and G. Grest, Phys. Rev. E {\bf 51},  5976  (1995).

\bibitem{Weis:93a}
J. Weis and D. Levesque, Phys. Rev. Lett. {\bf 71},  2729  (1993).

\bibitem{Stevens:95a}
M. Stevens and G. Grest, Phys. Rev. E {\bf 51},  5962  (1995).

\bibitem{Letter}
B. Groh and S. Dietrich, Phys. Rev. Lett. {\bf 72},  2422  (1994).

\bibitem{Paper}
B. Groh and S. Dietrich, Phys. Rev. E {\bf 50},  3814  (1994).

\bibitem{Klapp:97}
S. Klapp and F. Forstmann, J. Chem. Phys. {\bf 106},  9742  (1997).

\bibitem{Sear:96}
R. Sear, Phys. Rev. Lett. {\bf 76},  2310  (1996).

\bibitem{Osipov:96}
M. Osipov, P. Teixeira, and M.~T. da~Gama, Phys. Rev. E {\bf 54},  2597
  (1996).

\bibitem{Solid}
B. Groh and S. Dietrich, Phys. Rev. E {\bf 54},  1687  (1996).

\bibitem{Klapp:97a}
S. Klapp and F. Forstmann, Europhys. Lett. {\bf 38},  663  (1997).

\bibitem{Klapp:98}
S. Klapp and F. Forstmann, J. Chem. Phys. {\bf 109},  1062  (1998).

\bibitem{Klapp:00}
S. Klapp and G. Patey, J. Chem. Phys. {\bf 112},  10949  (2000).

\bibitem{Tao:91}
R. Tao and J. Sun, Phys. Rev. Lett. {\bf 67},  398  (1991).

\bibitem{Martin:98}
J. Martin, R. Anderson, and C. Tigges, J. Chem. Phys. {\bf 108},  3765  (1998).

\bibitem{Chen:92}
T. Chen, R. Zitter, and R. Tao, Phys. Rev. Lett. {\bf 68},  2555  (1992).

\bibitem{Dassanayake:00}
U. Dassanayake, S. Fraden, and A. van Blaaderen, J. Chem. Phys. {\bf 112},
  3851  (2000).

\bibitem{Gao:00}
G. Gao and X. Zeng, Phys. Rev. E {\bf 61},  2188  (2000).

\bibitem{Weis:93}
J. Weis and D. Levesque, Phys. Rev. E {\bf 48},  3728  (1993).

\bibitem{Lomba:98}
E. Lomba, J. Weis, and C. Tejero, Phys. Rev. E {\bf 58},  3426  (1998).

\bibitem{Fernandez:00}
J. Fern\'andez and J. Alonso, Phys. Rev. B {\bf 62},  53  (2000).

\bibitem{Domains}
B. Groh and S. Dietrich, Phys. Rev. E {\bf 53},  2509  (1996).

\bibitem{Simann}
B. Groh and S. Dietrich, Phys. Rev. E {\bf 57},  4535  (1998).

\bibitem{Smith:88}
A. Smith and N. Ashcroft, Phys. Rev. B {\bf 38},  12942  (1988).

\bibitem{Nijboer:57}
B. Nijboer and F. de~Wette, Physica {\bf 23},  309  (1957).

\bibitem{Slater:67}
J. Slater, {\em Insulators, Semiconductors, and Metals} (McGraw-Hill, New York,
  1967).

\bibitem{Denton:89}
A. Denton and N. Ashcroft, Phys. Rev. A {\bf 39},  4701  (1989).

\bibitem{Curtin:86}
W. Curtin and N. Ashcroft, Phys. Rev. Lett. {\bf 56},  2775  (1986).

\bibitem{Gay:98}
S. Gay, P. Beale, and J. Rainwater, J. Chem. Phys. {\bf 109},  6820  (1998).

\bibitem{Rosensweig:85}
R. Rosensweig, {\em Ferrohydrodynamics} (Cambridge University Press, Cambridge,
  1985).

\bibitem{ER:96}
{\em Electro-Rheological Fluids, Magneto-Rheological Suspensions and Associated
  Technology}, edited by W.~A. Bullough (World Scientific, Singapore, 1996).

\bibitem{Dubois:00}
E. Dubois, R. Perzynski, F. Bou\'e, and V. Cabuil, Langmuir {\bf 16},  5617
  (2000).

\bibitem{Lemaire:92}
E. Lemaire, G. Bossis, and Y. Grasselli, Langmuir {\bf 8},  2957  (1992).

\bibitem{Woestman:93}
J. Woestman and A. Widom, Phys. Rev. E {\bf 48},  1995  (1993).

\bibitem{Marshall:89}
L. Marshall, C. Zukoski, and J. Goodwin, J. Chem. Soc. Faraday Trans. I {\bf
  85},  2785  (1989).

\bibitem{Adams:76}
D. Adams and I. McDonald, Mol. Phys. {\bf 32},  931  (1976).

\bibitem{KittelEwald}
C. Kittel, {\em Einf\"uhrung in die Festk\"orperphysik}, 12th ed. (R.
  Oldenbourg, Munich, 1999).

\end{thebibliography}

%%%%%%%%%%%%%%%%%%%%%%%%%%%%%%%%%%%%%%%%%%%%%%%%%%%%%%%%%%%%%%%%%%%%%%

\newpage

\newpage

\end{document}